\documentclass[twocolumn,showpacs,preprintnumbers,amsmath,amssymb]{revtex4}

\usepackage{graphicx}
\usepackage{dcolumn}
\usepackage{bm}

\begin{document}
\preprint{ }

\title{Self-consistent Coulomb effects and charge distribution of quantum dot arrays}

\author{R. Wetzler\email{wetzler@physik.tu-berlin.de}, A. Wacker,
and E. Sch{\"o}ll}

\affiliation{Institut f{\"u}r Theoretische Physik,
Technische Universit{\"a}t Berlin, Hardenbergstr.~36, 10623~Berlin, Germany}
\date{\today}

\begin{abstract}
This paper considers  the self-consistent Coulomb interaction within  arrays
of self-assembled InAs quantum dots (QDs) which are embedded in a pn
structure. Strong emphasis is being put on the
statistical occupation of the electronic QD states which
has to be solved self-consistently with the actual three-dimensional potential
distribution.  
A model which is  based on a Green's function formalism including
screening effects  is used to calculate the interaction of QD carriers
within an array of QDs, where screening due to
the inhomogeneous bulk charge
distribution is taken into acount.  We apply our model to simulate capacitance-voltage (CV)
characteristics of a pn structure with embedded QDs.  Different size
distributions of QDs and ensembles of spatially perodic and randomly distributed arrays of QDs are investigated.
\end{abstract}
\pacs{73.63.Kv}

\maketitle

\section{Introduction}
The strong localization of few electrons in three dimensions is one of the key
features of quantum dots \cite{bim99} (QDs).  This results in a variety of
properties making them potential candidates for new semiconductor applications such 
as laser and memory devices \cite{gru00,fin98,yus97}.  One of those
properties is the singular density of states, which relies on the alignment of
the energy  levels in the whole QD ensemble. On the one hand this requires a
fairly homogeneous QD size distribution, on the  other hand the statistical
occupation of electronic QD levels \cite{gru97} changes the energy of levels of
the same dot and all neighboring QDs due to Coulomb interaction.

The amount of Coulomb repulsion between QD electrons, especially at different
QDs, is expected to depend on the details of the complicated bulk charge
distributions of the  QD device.  Additionally, the   microscopic charging and
discharging processes of the QDs depend on the energy differences
between the QD levels and the  non-equilibrium
quasi-Fermi level, resulting in a very complicated self-consistent problem.  The
microscopic charge distribution in the QD array is never stationary due to the statistical
character of microscopic  processes, but for large ensembles of QDs or in a
long time average a stationary mean QD sheet charge density is attained.  
Standard approaches use quasi one-dimensional models 
where the charge in the QDs is approximated by a homogeneous sheet charge
 \cite{bro98b,bru99}. This is suitable to obtain the QD energy
levels by a fit to experimental capacitance-voltage (CV) data \cite{wet00}, 
where a Gaussian distribution of  QD energy levels was assumed.
Nevertheless the  interpretation of the fitted broadenings  remains controversial, 
since they may result both from size fluctuations and inhomogeneous
charging effects of the single QDs.

The model used in this work calculates self-consistently the charge
distribution in an array of QDs, with respect to the three-dimensional
potential distribution under bias conditions.  From an ensemble average of
many QD charge configurations, the mean stationary QD sheet charge is
calculated, to obtain macroscopic quantities like  the depletion capacitance and the
CV characteristics.  The use of the  electrostatic Green's function method
makes this expensive self-consistent problem solvable and  charging effects in
the QD array can be investigated separately  from structural fluctuations.

\section{Theory}
Since only a small but essential part of the structure is not translationally
invariant in the lateral ($x$,$y$) directions, but most of the device is
inhomogeneous  only in the growth direction ($z$) the problem of solving the
three-dimensional non-linear Poisson's equation  is divided up into
calculating two one-dimensional potentials $\Phi$ and $\Phi^{\text{BQ}}$ and a
suitable 3D Green's function.  The potential $\Phi$ is introduced as the
self-consistent solution of the non-linear one-dimensional Poisson's equation
\begin{eqnarray}
\label{4poisson}
\epsilon_0\partial_z[\epsilon(z)\partial_z\Phi(z)]=-\rho(\Phi,E_{\text{Fn}},E_{\text{Fp}})-\sigma^{\text{QD}}
\chi_{\text{QD}}(z)
\end{eqnarray}
and the current equations neglecting generation and recombination in the active region 
\begin{equation}
\begin{split}
&\partial_zj_n[\Phi,E_{\text{Fn}}]=0\\
&\partial_zj_p[\Phi,E_{\text{Fp}}]=0\text{,}
\end{split}
\end{equation}
where the QD charge is treated as a
 sheet charge $\sigma^{\text{QD}}$  (per unit area) averaged over the lateral
 directions.
The bulk charge density $\rho$ is a function of the
local potential $\Phi$, and the electron and hole quasi-Fermi levels $E_{\text{Fn}}$ and $E_{\text{Fp}}$.
The electron current $j_n$ and the hole current $j_p$ are calculated within a drift-diffusion model.
Ohmic contacts are used to obtain boundary conditions for $\Phi$, $E_{\text{Fn}}$ and $E_{\text{Fp}}$.
For further details see  Ref. \cite{wet00}.
  Here, $\epsilon_0$ and $\epsilon$ denote the absolute and
 relative permittivities of the semiconductor material, and $\chi_{\text{QD}}$ is
 the characteristic function
\begin{eqnarray}
\chi_{\text{QD}}(z)&=&1/h\;\; \text{for}\;\;z\in \text{QD layer} \\ \nonumber
\chi_{\text{QD}}(z)&=&0\;\; \text{else,} \\ \nonumber
\end{eqnarray}
where $h$ is the height of the QDs, accounting for the localization in $z$
direction.

$\Phi^{\text{BQ}}$ is the potential which is felt by a single QD charge,
originating from the bulk charge if the QD charges would not exist.
$\Phi^{\text{BQ}}(z_{\text{QD}})$, where $z_{\text{QD}}$ is the position of
the QDs, can be conceived as   a background potential for the QDs.
$\Phi^{\text{BQ}}$ is calculated as $\Phi$ but without $\sigma^{\text{QD}}$
from eqn. (\ref{4poisson}).

In this paper the intrinsic QD level energy $E^{\text{intr}}_{\alpha \mu}$  of
a state  $\mu=(n_x,n_y,n_z)$ in QD $\alpha$ (where $n_x$, $n_y$, $n_z$ are the respective quantum numbers) is
measured from the conduction band edge.  In order to specifically  account for the size
dependence of the level energies in  InAs QDs we use a fit to
the results of $\boldsymbol{k\cdot p}$ calculations  \cite{sti99} as shown in
Fig. \ref{E0}.
 
The charging energy $E^{\text{char}}_{\alpha \mu}$   originating from the
Coulomb interaction of the QD electrons is treated as a first order
perturbation.  Thus the QD electron energy can be expressed  as
\begin{equation}
E^{\text{QD}}_{\alpha
\mu}=E_{\text{C0}}(z_{\text{QD}})-e\Phi^{\text{BQ}}(z_{\text{QD}})+E^{\text{intr}}_{\alpha
\mu}+E^{\text{char}}_{\alpha \mu}\text{.}
\end{equation}
Here $E_{\text{C0}}$  denotes the intrinsic conduction band edge, and $e>0$ is
the elementary charge.  The Coulomb charging energy $E^{\text{char}}_{\alpha
\mu}$  can be expressed by the use of the capacitance matrix $C$ \cite{wha96}
as
\begin{equation}
E^{\text{char}}_{\alpha
\mu}=e^2\sum_{\beta\nu}p_{\beta\nu}\left(C^{-1}\right)_{\alpha\beta}\left(1-\delta_{\alpha
\mu,\beta \nu}\right)\text{.}
\end{equation}
Here $p_{\beta\nu}$ denotes the occupation numbers of the QD state $\nu$ in QD
 $\beta$ ($p_{\beta\nu}=1$ if occupied, $p_{\beta\nu}=0$ if not
occupied). $\delta_{\alpha \mu,\beta \nu}$  ensures that an electron does not
interact with itself.  The charging energies obtained by summing over
$\beta=\alpha$ will be referred to as intradot charging energy, where the  sum
over $\beta\ne\alpha$ is referred to as interdot charging energy in this
paper.  The elements of the inverse of the capacitance matrix
$\left(C^{-1}\right)_{\alpha\beta}^{\mu\nu}$ can be expressed by the use of the
Green's function  $G$ \cite{war98}:
\begin{equation}
\left(C^{-1}\right)_{\alpha\beta}^{\mu\nu} \equiv\int\limits_{V}\int\limits_{V}
\left|\psi_{\alpha\mu}(\boldsymbol{r})\right|^2 \left|\psi_{\beta\nu}(\boldsymbol{r}')\right|^2G(\boldsymbol{r},\boldsymbol{r}')  d^3\boldsymbol{r}'  d^3\boldsymbol{r}
\end{equation} 
Here $\psi_{\alpha\mu}$ and $\psi_{\beta\nu}$ symbolize the QD electron wavefunctions.
Assuming that the QD electrons are distributed homogeneously
 within cylinder shaped QDs with height $h$ and a diameter $D_{\alpha}$,
 neglecting the details of the QD wave functions leads to 
\begin{equation}
\left(C^{-1}\right)_{\alpha\beta
}=\frac{1}{V_{\alpha}V_{\beta}}\int\limits_{V_{\alpha}} \int\limits_{V_{\beta}}
G(\boldsymbol{r},\boldsymbol{r}')  d^3\boldsymbol{r}'  d^3\boldsymbol{r}\text{,}
\end{equation} 
where $V_{\alpha}$  and $V_{\beta}$ denote the volumes of two QDs $\alpha$ and
 $\beta$.  In our model the
 charging energies are the same for all states $\mu,\nu$ within a particular QD.  A
 Green's  function $G$ including  the  static  screening effects by free bulk
 charges can be found by defining a screening parameter $f_{\text{scr}}$ as
\begin{equation}
f_{\text{scr}}(z) \equiv\frac{\partial\rho}{\partial\Phi}(\Phi^{\text{BQ}}(z))\text{.}
\end{equation}
Then $G$ is calculated as \cite{lan91}
\begin{eqnarray}
\label{green}
& &\left[\Delta_{\boldsymbol{r}}+\partial_z\left(\ln\epsilon(z)\partial_z\right)+\frac{f_{\text{scr}}(z)}{\epsilon(z)\epsilon_0}\right]G(z,r,z')
\\ \nonumber & &  =
-\frac{1}{\epsilon(z)\epsilon_0}\delta((0,z')-(r,z))\text{,}
\end{eqnarray} 
where $\Delta_{\boldsymbol{r}}$ is the Laplace operator  and $r$ is the radial distance in the
$x-y$ plane.  The solution of eqn.(\ref{green}) has a cylindrical symmetry, since
$f_{\text{scr}}$ and $\epsilon$ depend only on $z$.  Note that $G$ is the
potential of a unit point-like charge positioned at $z'$, $r'=0$.

The QDs are occupied statistically according to the level energies
$E^{\text{QD}}_{\alpha \mu}$ and the contact quasi-Fermi level
$E^c_{\text{Fn}}$ with the occupation probability
\begin{equation}
f=\frac{1}{1+\exp\left(\frac{E^{\text{QD}}_{\alpha
\mu}-E^c_{\text{Fn}}}{k_BT}\right)}\text{,}
\end{equation}
where $T$ is the temperature and $k_B$ denotes Boltzmann's constant.  Note
that every change in the occupation numbers $p_{\alpha\mu}$ changes all QD
level energies.  The stationary, mean QD sheet charge density
$\sigma^{\text{QD}}$ used in eqn. (\ref{4poisson}) is found as a charge
configuration  average
\begin{equation}
\sigma^{\text{QD}}=\frac{-e}{A}\frac{1}{N_i}\sum^{N_i}_{i=1}
\sum_{\alpha,\mu}p^i_{\alpha\mu}\text{,}
\end{equation}
where $i$ denotes $N_i$ simulated realizations of configurations of occupied QD levels, and
$A$ is the quadratic contact cross section simulated.  Periodic boundary conditions are used in the 
four parallel and four diagonal directions  in  the QD plane.
Since  $\sqrt{A}$, which
takes values from 1~$\mu$m to 3~$\mu$m in our simulations, is always larger than
the maximum expected screening length for this structure, the influence
of boundaries is negligible. In other words, a QD charge does not ``see''
itself in the periodically continued plane.  The stationary depletion layer
capacitance is determined as in Ref. \cite{wet00} by the electric field in the
depletion layer given by  $\Phi$.

\section{Results}
\begin{figure}
\includegraphics[angle=0,width=1.0\columnwidth]{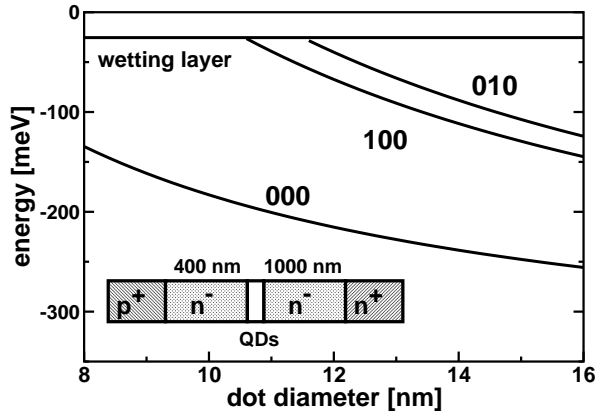}
\caption[a]{Size dependence of the electron levels in InAs QDs, according to a fit to $\boldsymbol{k\cdot p}$ calculations \cite{sti99}.
The QD levels are labelled by the quantum numbers $(n_x,n_y,n_z)$. Inset: The pn structure under investigation with embedded self-organized quantum dots. }
\label{E0}
\end{figure}

The pn structure considered here consists of highly doped $n^+$ and $p^+$ GaAs
contact regions, as well as a $3 \times 10^{16} cm^{-3}$ $n^-$-doped GaAs
region.  A layer of self-organized InAs QDs  is positioned 400 nm from the pn
interface. The structure is displayed in the inset of Fig. \ref{E0}.  Similar
pn structures have been investigated experimentally by CV spectroscopy and
DLTS measurements \cite{kap99,kap00a}.    We simulate arrays of up to $30$ by
$30$ QDs with a QD sheet density of $N_{QD}=10^{10}cm^{-2}$ for  various
temperatures and ensembles of QDs. The average QD diameter is 12~nm. A QD
height of 3 nm is used.

\begin{figure}
\includegraphics[angle=-90,width=1.0\columnwidth]{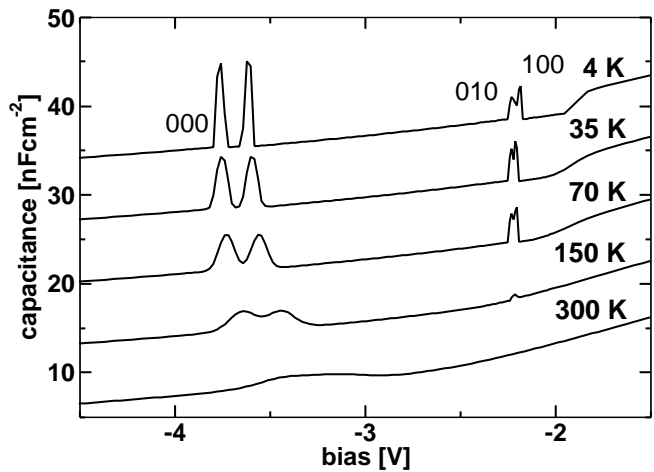}
\caption[a]{Calculated capacitance-voltage characteristics of the  pn diode under investigation with periodically ordered and equally sized InAs quantum dots with a diameter of 12 nm and a height of 3 nm for various temperatures. Plots for 300~K, 150~K, 35~K and 4~K are offset by -14, -7, +7 and +14~$\text{nF cm}^{\text{-2}}$ for clarity.}
\label{tempdep}
\end{figure}

\subsection{Ideal dot distribution}
When a reverse bias $U$ is applied to the diode, bulk carriers as well as QD
states will be successively discharged \cite{wet00}.  First we simulate the
structure at 4~K, 35~K, 70~K, 150~K and 300~K  with periodically ordered QDs
and  with all QDs having  the same size.  The resulting CV characteristics are
shown in Fig. \ref{tempdep}.  The main drop of the capacitance with
increasing reverse bias is due to the increasing bulk depletion region
\cite{sze81}. On top of this background
three peaks can clearly be observed for temperatures
less than 150~K.  The peak at $U\approx$-2.2~V originates from the discharging
of the excited states, which is split since piezoelectric fields break the
degeneracy of the states (100) and (010) (see Fig. \ref{E0}).  At
$U\approx$-3.6~V the two ground state electrons are discharged successively
leading to a double peak.  Here, the energy degeneracy is broken by the
Coulomb charging effects.  The Coulomb splitting has also been observed
experimentally for different structures \cite{med97,med97a,luy99}.  The peak
structure is broadened with rising temperature, resulting in a plateau in the
CV characteristics at 300~K.  Our  calculations  use  a   charging energy 
of about 20~meV.  From the ground state peak splitting  in  the 4~K curve of
about 100~mV we can estimate a lever-arm factor of 200~meV/V which, when applied
to the width of the ground state peaks, results in an energy broadening of
2.5~meV. This value coincides with  the unscreened Coulomb potential (in units
of energy)  of two electrons at a distance of 100~nm  from a QD
(2~$\times$~1.25~meV).  This means that the broadening at 4~K originates from the
charge fluctuations within  the four  nearest neighbor QDs, in particular if every
second one is charged. At 300 K, where occupation is rather insensitive to the QD level spacing, 
the full plateau width of about 500~mV corresponds to an energy broadening of 100~meV.
This value  can be attributed to the broadening in InAs QDs of the Fermi distribution function of about $2$ $k_BT$ at 300~K 
 of the two ground state levels.  
Consequently, at large temperatures, the charging energy is given by the
averaged, statistically distributed charge stored in the QD array.
This is the reason why the CV
characteristics shows a flat plateau at 300~K and  is
hardly  distinguishable from that of a quantum well \cite{wan96}.

Additionally, the ground state peaks show a shift to lower reverse bias for
rising temperatures, where the excited states peaks do not show this feature.
This shift can be explained as follows: When the ground states electrons are
discharged, interdot charging energies become more important than for the
excited states (explained later in detail).  The interdot interaction depends
on how the free electrons can screen the interdot Coulomb potential. 
 This  ability rises with lower temperatures leading to
shorter screening lengths.  Therefore the charging energies are lower at low
temperatures, and a higher reverse bias has to be applied to discharge the
ground state electrons.

\subsection{Impact of structural fluctuations}
\begin{figure}
\includegraphics[angle=-90,width=1.0\columnwidth]{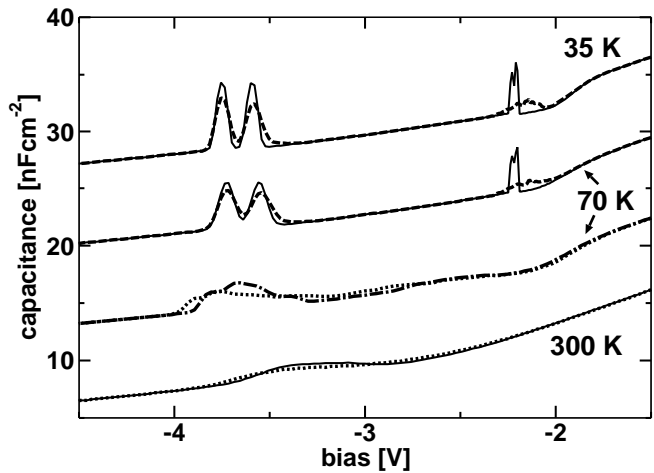}
\caption[a]{Calculated capacitance-voltage characteristics at 35~K, 70~K and 300~K without structural quantum dot fluctuations (solid lines), 
with randomly distributed  but equally sized dots (dashed lines), with size
fluctuations of $\pm$17~\% (dash-dotted line) and $\pm$33~\% (dotted
lines). The curves for 35~K, 70~K with size fluctuations and 300~K are offset
by +7, -7 and -14~$\text{nF cm}^{\text{-2}}$.}
\label{CV}
\end{figure}

 Now we investigate the impact of  structural fluctuations of the QD ensemble
such as position disorder and size
fluctuations.  At first we distribute the equally
sized QDs randomly instead of periodically in the lateral plane leaving
the QD sheet density constant.  Compared to the characteristics without
fluctuations (solid lines in Fig.~\ref{CV}) at 35~K and 70~K the ground state
peaks   for randomly distributed QDs are  only slightly smeared out (dashed
lines in Fig.~\ref{CV}).  Since the electrons in the excited states can be
thermally activated  more easily due to fluctuations in the interdot Coulomb
interaction, resulting in an inhomogeneous charging of these levels, the
respective peaks are more strongly broadened.
 
Next we simulate  an ensemble of periodically ordered QDs with random
QD diameters  between 10 and 14~nm at 70~K,  corresponding to a fluctuation of
{$\pm$ 17~\%}.  The resulting CV characteristic (see dash-dotted line in
Fig.~\ref{CV}) shows a very broad ground state peak.  Again, the excited
states are more sensitive to the fluctuations, i.e., the respective peak disappears almost completely.
  The increase of the fluctuations   of the diameter
to $\pm$33~\% (8 - 16~nm) leads to a broad ground state plateau in the
characteristics (dotted  lines in Fig.~\ref{CV}).  Such a plateau has been
observed experimentally for similar pn structures \cite{kap99,kap00a},  and a
 comparison with our results indicates that the size fluctuations
in self-organized QDs is of  this order of magnitude. 
This size fluctuation
corresponds to  an energy fluctuation of the ground state of about 100~meV
(Fig.~\ref{E0}), which coincides with the typical value for the broadening of
the ground state in the photoluminescence signal of InAs QDs \cite{kap99}.  For room
temperature the CV characteristics with $\pm$33~\% size fluctuation shows almost no
deviation from the characteristics  without  fluctuations. Here, thermal charge
fluctuations  dominate the behavior.

\begin{figure}
\includegraphics[angle=0,width=1.0\columnwidth]{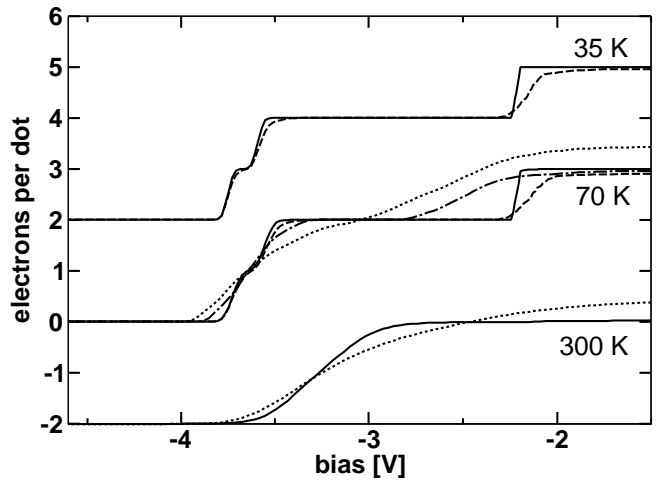}
\caption[a]{Mean number of electrons per QD as a function of bias for 35~K,
  70~K and 300~K without structural quantum dot fluctuations (solid lines),
  with randomly distributed but equally sized dots (dashed lines), with size
  fluctuations of $\pm$17~\% (dash-dotted line) and $\pm$33~\% (dotted
  lines). The curves for 35~K, and 300~K are offset by +2 and -2 electrons per
  dot, respectively.}
\label{dens}
\end{figure}

As already mentioned, the QDs are discharged by applying a reverse bias.  In
Fig.~\ref{dens} the mean number of electrons per QD is displayed as a function
of bias for 35~K, 70~K and 300~K.  At 35~K and 70~K the QDs are charged with
three electrons at zero bias and are discharged monotonically in a  step-like manner at
biases where peaks are observed in the CV characteristics (solid lines).
Structural fluctuations lead to a broadening of these curves (dashed,
dash-dotted and dotted lines).  Additionally, with size fluctuations of $\pm$33~\%
some QDs can store up to four electrons, leading to a higher average QD
occupation.  For 300~K the QDs are charged with less electrons, because of the
lower degree of degeneracy of bulk  electrons in the contact region, and
consequently the  lower lying quasi-Fermi level at higher temperatures.

\subsection{Variation of the charging energy}
\begin{figure}
\includegraphics[angle=0,width=1.0\columnwidth]{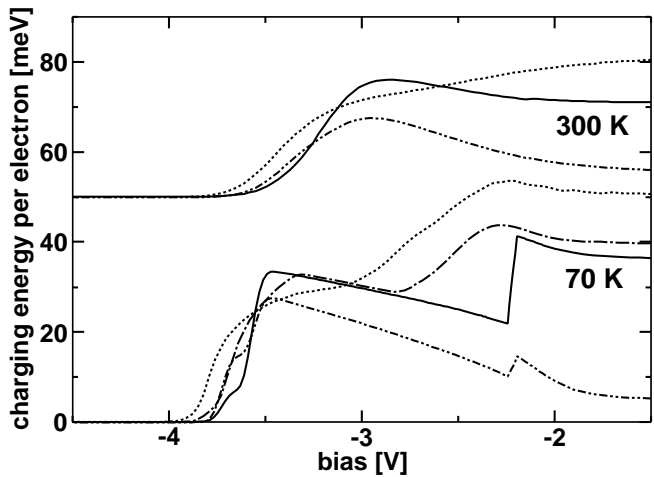}
\caption[a]{The mean charging energy per QD electron for 70~K and 300~K without structural quantum dot fluctuations (solid lines), with randomly distributed  but equally sized dots (dashed lines), with size fluctuations of $\pm$17~\% (dash-dotted line) and $\pm$33~\% (dotted lines) as calculated from the 3D simulations. The average potential barriers as calculated from the 1D Poisson's equation are also shown (double dot-dashed  line). The curves for 300~K are offset by +50~meV.   }
\label{charging}
\end{figure}

The mean charging energy per QD electron as shown in Fig.~\ref{charging} shows
an interesting, non-monotonic behavior.  At 70~K  without fluctuations
(solid lines) the charging energy is at first constant for $U<$-0.5~V and then
rises until the  excited state electrons are rapidly discharged. Between
$U$=-2.2~V and $U$=-3.6~V the charging energy again rises, and then  drops to
zero when the residual ground state electrons are discharged.  The continuous
increase of the charging energies can be explained by considering the bulk
electrons in the vicinity of the QDs being depleted with reverse bias. This
depletion increases  the interdot charging energy, and therefore the total QD
level energies.  Fig.~\ref{charging} shows that these  negative and positive
differential charging energies persist even  for  $\pm$33~\% size fluctuations at
70~K. Also for 300~K this behavior can be observed, but it vanishes when
taking into account additionally $\pm$33~\% size fluctuations.  The mean height of
the potential  barrier
$e[\Phi(z_{\text{QD}})-\Phi^{\text{BQ}}(z_{\text{QD}})]$ at the QD sheet as
obtained from the one-dimensional Poisson's equation is also displayed in
Fig.~\ref{charging}.  It shows the big difference of the charging energies
between one-dimensional and three-dimensional calculations, especially if the
QDs are charged with more than two electrons.  But nevertheless, it shows the
same differential behavior, which has an inhibitory effect on
leakage and recombinations currents in QD devices, and therefore may provide an
explanation for the observed bistabilities in similar pn structures \cite{kie03}.
  
\begin{figure}
\includegraphics[angle=0,width=1.0\columnwidth]{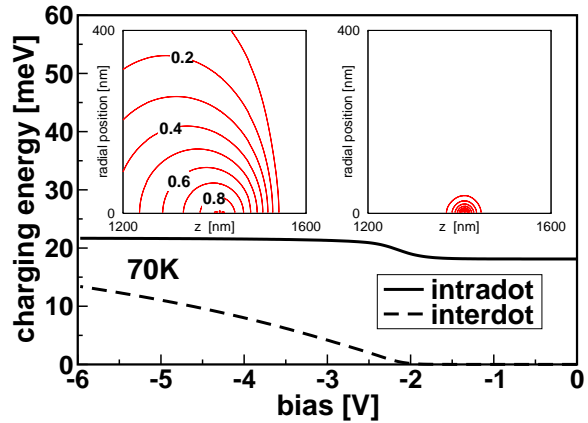}
\caption[a]{The intradot (solid line) and interdot charging energy (dashed line) as a function of bias at 70~K. Insets: The Green's functions (normalized to the unscreened Green's function) at $U$=-1.5~V (left) and $U$=-4.5~V (right) visualize the effect of the depletion of the free electron gas upon the charging energies.}
\label{inter}
\end{figure}

To explain the origin of the negative differential charging energies in more
detail, the charging energies are plotted in Fig.~\ref{inter} as a function of
bias, assuming  that all QDs are charged with one electron at any  bias and adding
one extra electron into a single QD.  We also differentiate between the intradot 
charging energy (solid line) originating from the  second electron in  a
QD, and the interdot charging energy (dashed line) from the Coulomb repulsion
with all neighboring QDs. The ratio of the Green's function and the unscreened
Green's function ($f_{\text{scr}}=0$)  is also displayed in the  two  insets
of Fig.~\ref{inter} for $U$=-1.5~V and $U$=-4.5~V as a function of the lateral
position $r$ and the vertical position $z$ as contour plots.  As shown in Section II, the
Green's function determines the capacitance matrix and therefore the charging
energies. As the Green's function spreads out more and more, especially in the lateral
directions, the charging energies due to interdot charging
effects become larger.  The effect on the intradot charging energy is rather small, since
the size of the QDs is small compared to the distance of the free electron gas
and the respective screening length.  But by depletion of  the free electrons with
the applied reverse bias, the interdot charging rises.  The slope  of the
interdot charging energy can clearly be identified with the slope of the mean
charging energy of  Fig.~\ref{charging} for biases where  the QD charge is
constant but non-zero.

\begin{figure}
\includegraphics[angle=0,width=1.0\columnwidth]{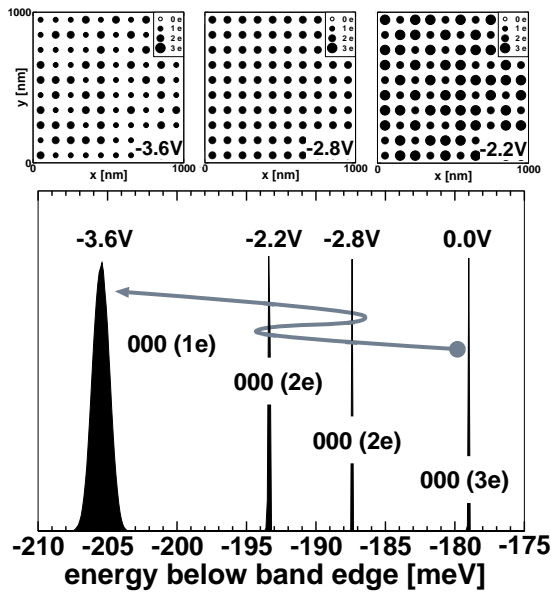}
\caption[a]{The electron distribution within the quantum dot layer (top)  and the lineshape of the ground state (000) for different reverse biases at 70~K (bottom).
 The  discharging of the dots and the bias dependent Coulomb interaction lead
 to a nonmonotonic shift of the energy distribution.  The charge inhomogeneity in the
 quantum dot layer leads to a broadening of the energy distribution at
 $U$=-3.6~V. }
\label{densline}
\end{figure}

Fig. \ref{densline} shows the QD charge distribution within  the QD layer  for
$U$=-2.2~V, $U$=-2.8~V and $U$=-3.6~V  and the  energy distribution of the
ground state (000) at  $U$=0.0~V,  $U$=-2.2~V, $U$=-2.8~V and $U$=-3.6~V for
70~K and without structural fluctuations. The normalizations of the
distributions have been chosen differently for clarity.  The QD charge
distribution is inhomogeneous when the QDs are discharged at  $U$=-2.2~V and
$U$=-3.6~V, but  it is homogeneous at $U$=-2.8~V.  Fig.~\ref{densline} shows
that the initial ground state energy at $U$=0.0~V  is lowered due to
discharging one electron at $U\simeq$-2.2~V.  Then the energy rises (up to
$U\simeq$-2.8~V), due to the rising interdot charging energies as argued
above.  The discharging of the ground state electron then leads to a lowering
of the QD electron energies again. Overall, the QD energy shows a zig-zag
behavior when sweeping  the bias.

The  inhomogeneity of the QD charge leads also  to a line broadening of the
electron states.  For $U$=-2.8~V and $U$=0.0~V  the charge distribution is
homogeneous, and the broadening vanishes.  As shown above the magnitude of the
Coulomb interaction between the QDs depends on the bias (Fig.~\ref{inter}).
Therefore the line broadening at $U$=-2.2~V is still small even for an
inhomogeneous QD charge, since the interdot interaction is close to zero.
The broadening becomes larger (about 2~meV) at $U$=-3.6~V, since at this bias
the charge distribution again is inhomogeneous, and the interdot interaction
is large (see Fig.~\ref{inter}).

\section{Conclusion}
In conclusion, we have proposed a model to calculate the self-consistent
three-dimensional charge  distributions and charging energies within a QD
device.  We have shown that the charging energies depend on the electronic
structure, which is determined by doping concentration, bias and temperature.
The effects of size fluctuations of the dots and the role of a spatially periodic or
 random distribution of the dots on the capacitance-voltage characteristics
have been investigated. We have found that size fluctuations have a much
stronger influence on the broadening of the peak structure of the CV
characteristics than the positional disorder of the QDs.
Our detailed simulations have shown that the QD
energy can be a strongly non-monotonic function of the applied bias.

\begin{center}
{\bf Acknowledgement}\\
\end{center}
This work was supported by DFG in the framework of Sfb 296. 
Helpful discussions with C. Kapteyn and R. Heitz are acknowledged.

\bibliographystyle{prsty}

\end{document}